\documentclass[prl,showpacs,twocolumn,aps,superscriptaddress,a4paper]{revtex4-1}
\usepackage{dcolumn,amssymb,amsmath,amsfonts,graphicx,latexsym,color}

\begin{document}

\title{Topological phase transition in the quench dynamics of a one-dimensional Fermi gas}
\author{Pei Wang}
\affiliation{Department of Physics, Zhejiang University of Technology, Hangzhou 310023, China}
\affiliation{International Center for Quantum Materials, School of Physics, Peking University, Beijing 100871, China}
\author{Wei Yi}
\affiliation{Key Laboratory of Quantum Information, University of Science and Technology of China, CAS, Hefei, Anhui, 230026, People's Republic of China}
\affiliation{Synergetic Innovation Center of Quantum Information and Quantum Physics, University of Science and Technology of China, Hefei, Anhui 230026, China}
\author{Gao Xianlong }
\affiliation{Department of Physics, Zhejiang Normal University, Jinhua 321004, China}

\date{\today}

\begin{abstract}
We study the quench dynamics of a one-dimensional ultracold Fermi gas in an optical lattice potential with synthetic spin-orbit coupling. At equilibrium, the ground state of the system can undergo a topological phase transition and become a topological superfluid with Majorana edge states. As the interaction is quenched near the topological phase boundary, we identify an interesting dynamical phase transition of the quenched state in the long-time limit, characterized by an abrupt change of the pairing gap at a critical quenched interaction strength. We further demonstrate the topological nature of this dynamical phase transition from edge-state analysis of the quenched states. Our findings provide interesting clues for the understanding of topological phase transitions in dynamical processes, and can be useful for the dynamical detection of Majorana edge states in corresponding systems.
\end{abstract}

\maketitle

\emph{Introduction}.--
Topological phases and phase transitions~\cite{wen90} in quantum many-body systems have recently attracted much attention in various physical contexts. A particularly interesting topological state is the topological superfluid (TSF) phase, where Majorana edge states may exist at its boundary~\cite{read00,kitaev00}. Besides systems with intrinsic chiral $p$-wave pairing superfluidity~\cite{pwave1,pwave2,pwave3,pwave4,gurarie}, TSF phases can also be induced from an $s$-wave pairing superfluid in the presence of spin-orbit coupling (SOC) and an effective Zeeman field~\cite{swave1,swave2}. While the search for Majorana zero modes is being actively pursued in condensed matter systems such as the semiconductor/superconductor heterostructures~\cite{fu08,wirereview}, the possibility of realizing topological superfluidity and hence Majorana zero modes in ultracold atomic gases has also been extensively studied, due to the recently implemented synthetic spin-orbit coupling in these systems~\cite{lin11,fermisocexp1,
fermisocexp2,chuanweisoc,zhoujingsoc}.

An advantage of ultracold atomic gases is the controllability, which not only allows for experimental characterization of the system over a wide range of parameters, but also provides a convenient way to study the dynamical processes. In recent years, dynamical processes in ultracold atomic gases play an increasingly important role in revealing key properties of the system~\cite{polkovnikov}. Particularly, quench dynamics, in which the system evolves from a pure state following a sudden change of parameters, has been widely used in cold atoms experiments~\cite{bloch,coldatomquench1,coldatomquench2,coldatomquench3,quench4,quench5}. In the long-time limit, the quenched state is typically quite different from an equilibrium state described by the ensemble theory~\cite{kollath,manmana}. More importantly, it has been shown that depending on the quench parameters, the quenched state in this limit can exhibit different dynamical phases~\cite{fosterPRB,fosterarxiv}.

With the prospect of realizing topological orders in ultracold atomic gases, it is timely to study dynamical processes in systems with topological order, and ask questions like: whether topological properties are robust against dynamical perturbations; how topology should be defined and probed in dynamical processes, etc. On the one hand, in an equilibrium state with topological order, features like edge states are protected by a bulk gap~\cite{nayak08}, while such protection may not be effective in dynamical processes, especially in quench dynamics, where the sudden change inevitably induces high-energy excitations~\cite{polkovnikov}. On the other hand, despite many recent studies on topological phases and edge states in quantum quenches, topology in a dynamical process is still an on-going research that needs further characterization \cite{fosterPRB,fosterarxiv,topoquench1,topoquench2,topoquench3,topoquench4,topoquench5}.

In this work, we theoretically study the quench dynamics of a one-dimensional ultracold Fermi gas in an optical lattice potential under the synthetic spin-orbit coupling that has recently been realized experimentally~\cite{lin11,fermisocexp1,fermisocexp2}. Starting from a ground state of the system, we implement a sudden change of the interaction strength and focus on the dynamics of the pairing gap as the system evolves under the quenched Hamiltonian. At sufficiently long evolution times, the pairing gap either approaches a finite value or oscillates in time, depending on the quench parameters. Interestingly, we identify a critical interaction strength in the quenched Hamiltonian, at which the pairing gap in the long-time limit undergoes an abrupt jump across the critical value that corresponds to the emergence of the topological phase transition in the ground state of the pre-quench Hamiltonian. We associate this critical interaction strength with a dynamical phase transition. We further demonstrate that
edge states exist in the long-time limit only for processes with quenched interaction strength below the critical value of the dynamical phase transition, which suggests that the dynamical phase transition is also of topological nature. The topological property of the quenched state is thus determined by the quenched Hamiltonian, regardless of the initial state. Our findings provide clues for the understanding of topological phase transitions in dynamical processes, and may have interesting implications for future experiments.

\begin{figure}[tbp]
\includegraphics[width=1.0\linewidth]{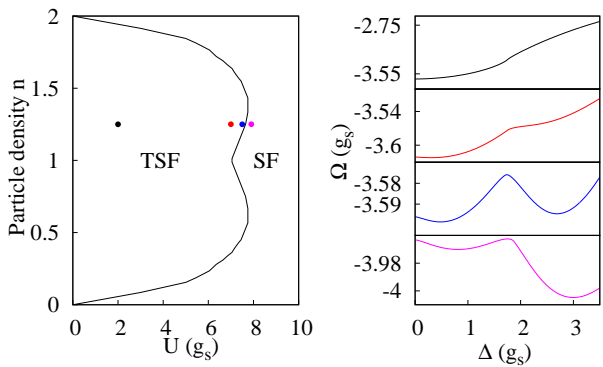}
\caption{(Color online) The left panel shows the equilibrium phase diagram with $x$-axis denoting the on-site interaction and $y$-axis the on-site occupation. The right four panels (from up to down) show respectively the thermodynamic potential as a function of $\Delta$ at four different points (from left to right) in the phase diagram.}\label{phasediagram}
\end{figure}

\emph{Model Hamiltonian and phases at equilibrium}.--
Under the single-band approximation, the tight-binding Hamiltonian can be written as:
\begin{align}\label{originalH}
 \hat H = & -g_s\sum_{j\sigma} (\hat c^\dag_{j\sigma} \hat c_{j+1,\sigma} +h.c.) + h\sum_j(\hat c^\dag_{j\uparrow} \hat c_{j\uparrow} - \hat c^\dag_{j\downarrow} \hat c_{j\downarrow}) \nonumber\\ & + \alpha \sum_j(\hat c^\dag_{j\uparrow} \hat c_{j+1,\downarrow} - \hat c^\dag_{j\downarrow} \hat c_{j+1,\uparrow} + h.c.) \nonumber\\ & - U \sum_j \hat c^\dag_{j\uparrow} \hat c_{j\uparrow} \hat c^\dag_{j\downarrow} \hat c_{j\downarrow},
\end{align}
where $\hat c^{\dag}_{j\sigma}$ is the creation operator of fermions on site $j$ with spin $\sigma=\uparrow,\downarrow$, and the on-site interaction is attractive with $U>0$. The effective Zeeman field $h$ and the spin-orbit coupling strength $\alpha$ are related to parameters of the Raman process generating the synthetic spin-orbit coupling.  

Following the stand BCS-type mean-field theory, we can Fourier-transform Hamiltonian (\ref{originalH}) into momentum space, and write the effective mean-field Hamiltonian in the grand canonical ensemble:
\begin{align}\label{effHmomen}
 \hat H_{eff} = & \sum_{k\sigma} \epsilon_{k\sigma} \hat c^\dag_{k\sigma} \hat c_{k\sigma} - \sum_k \left(\Delta \hat c^\dag_{k\uparrow} \hat c^\dag_{-k\downarrow} +h.c. \right) \nonumber \\ & + \sum_k (\alpha_k \hat c^\dag_{k\uparrow} \hat c_{k\downarrow} + h.c.)-\frac{|\Delta|^2}{U},
\end{align}
where $\hat c^\dag_{k\sigma}$ is the creation operator for fermions of momentum $k$ and spin $\sigma$, with $k$ defined on the first Brillouine zone. The dispersion of the lowest band $\epsilon_{k,\uparrow/\downarrow} = -2\cos k -\mu \pm h $, the spin-orbit coupling $\alpha_k = 2i\alpha \sin k$, and the pairing gap $ \Delta = U\sum_{k} \langle \hat c_{-k \downarrow} \hat c_{k\uparrow} \rangle$. The chemical potential $\mu$ is related to the on-site particle density $n$, which is kept fixed throughout the lattice. Here, we have $0<n<1$ for $\mu<0$, and $1<n<2$ for $\mu>0$. Due to the particle-hole symmetry of the effective Hamiltonian, we will focus on the case of $\mu>0$ henceforth.

At equilibrium, the ground state of the system can be determined by minimizing the thermodynamic potential, which at zero temperature is reduced to $\Omega=\langle \hat H_{eff}\rangle$. Here $\langle\cdots\rangle$ is taken with respect to the ground state. Previous studies show that the ground state of $\hat H_{eff}$ is a TSF with Majorana edge states for $\sqrt{(\mu - 2g_s)^2+ \Delta^2} \leq h \leq \sqrt{(\mu + 2g_s)^2+ \Delta^2}$, and is a conventional superfluid otherwise~\cite{xiaosen}. In Fig.~\ref{phasediagram}, we map out the ground state phase diagram at zero temperature, which serves as a basis for further studies of quench dynamics. Consistent with previous works, the system is in the TSF phase in the weakly-interacting regime, and undergoes a topological phase transition to become a conventional superfluid (SF) phase in the strongly-interacting regime with the phase boundary labeled by $U_c$. With our choice of parameters ($h=2g_s$ and $\alpha=g_s$), the transition boundary in Fig.~\
ref{phasediagram} is of first-order. This is manifested not only in the discontinuity of the pairing gap $\Delta$ at the transition boundary, but also in the shape of the thermodynamic potential $\Omega(\Delta)$ (see right panel of Fig.~\ref{phasediagram}). For the convenience of later discussions, we label the critical pairing gap of the topological phase transition $\Delta_c$, where $\Delta_c=\sqrt{h^2-(\mu-2g_s)^2}$.

\begin{figure}[tbp]
\includegraphics[width=1.0\linewidth]{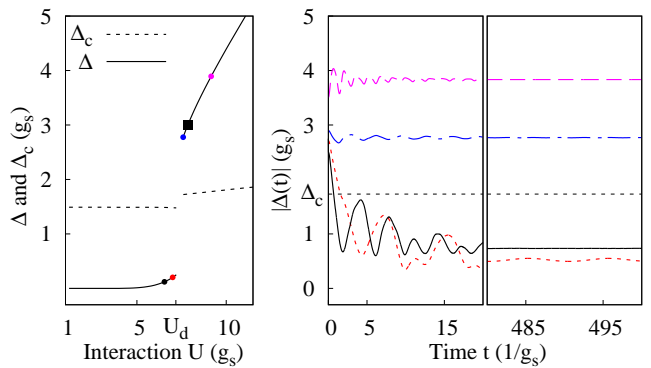}
\caption{(Color online) Quench dynamics with non-topological initial states. [Left panel] The pairing gap $\Delta$ in the ground state as a function of $U$ at $n=1.19$, $h=2g_s$ and $\alpha=g_s$. The discontinuity of $\Delta$ indicates the first-order phase boundary. The black square represents the initial state at $U_i$ and the circles of different colors represent the equilibrium pairing gap at $U_f$. [Right panel] The dynamics of $|\Delta(t)|$ for different $U_f$ marked in the left panel, shown by lines of the same color. The black dashed line shows the position of the critical $\Delta_c$ for the topological transition of the pre-quench state with $U_i$.}\label{fig:evolutiondelta3}
\end{figure}

\emph{Quench dynamics}.--
We now study the quench dynamics of the system where the interaction $U$ undergoes a sudden change. We assume that the system is initially in the zero-temperature ground state with $U=U_i$. At the start of the quench ($t=0$), the interaction is suddenly switched to $U=U_f$, driving the system out of equilibrium. The system then evolves under the Hamiltonian with the quenched interaction $U_f$.

To characterize the dynamics of the system, we introduce operators: $\hat \nu_{k\sigma} = \hat c^\dag_{k\sigma} \hat c^\dag_{-k\sigma}$, $\hat \tau_k=  \hat c^\dag_{k\uparrow} \hat c_{-k\downarrow}^\dag $, and $\hat \sigma_k= \hat c^\dag_{k\uparrow} \hat c_{k\downarrow}$. Together with the momentum-space density operator $\hat n_{k\sigma}=\hat c^\dag_{k\sigma} \hat c_{k\sigma}$, we may express the time-dependent mean-field effective Hamiltonian as:
\begin{align}
\hat H_{eff}(t) =&\sum_{k}\epsilon_k\left(\hat \tau^z_{k}+1\right)+\sum_{k}\left(\alpha_k \hat \sigma_{k} +h.c.\right) \nonumber\\ & -\sum_{k}\left(\Delta(t) \hat \tau_{k}+h.c.\right) + h\sum_{k}\hat \sigma^z_{k}-\frac{|\Delta|^2}{U},
\end{align}
where $\hat \tau^z_k=\hat n_{k\uparrow}+\hat n_{-k\downarrow}-1$, $\hat \sigma^z_k=\hat n_{k\uparrow}- \hat n_{-k\downarrow}$ and $\epsilon_k= -2\cos k -\mu$. From the Schr\"odinger's equations, one may derive a set of closed equations for the expectation values of these operators. The dynamical parameters of the system can be calculated numerically from these equations~\cite{supp}. Note that similar approach has been applied to study the quench dynamics in a Fermi gas throughout the BCS-BEC crossover without SOC \cite{levitov1,levitov2,burnett,duan}, and more recently, the quench dynamics of a chiral $p$-wave superfluid near the topological phase transition~\cite{fosterPRB}. With this approach, we focus on the dynamics of the pairing gap $\Delta(t)=U_f\sum_{k} \langle \hat c_{-k \downarrow} \hat c_{k\uparrow}\rangle$ during the evolution. Interestingly, the dynamics and the quenched states are quite different depending on the topological nature of the initial state.

\begin{figure}[tbp]
\includegraphics[width=1.0\linewidth]{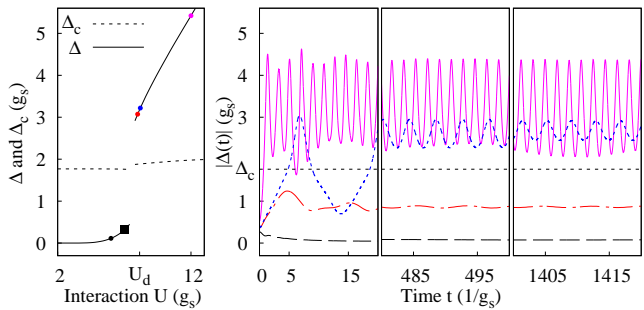}
\caption{(Color online)  Quench dynamics with topological initial states. [Left panel] $\Delta$ and $\Delta_c$ as a function of $U$ at equilibrium for $n=1.25$, $h=2g_s$ and $\alpha=g_s$. The black square represents the initial state at $U_i$ which is a topological superfluid, and the circles of different colors represent the equilibrium states at $U_f$. [Right panel] The dynamics of $|\Delta(t)|$ for different $U_f$ marked in the left panel, shown by lines of the same color.}\label{fig:evolutiondelta1}
\end{figure}

The right panel of Fig.~\ref{fig:evolutiondelta3} shows typical quench dynamics of the pairing gap when the initial ground state is a conventional superfluid, while the left panel illustrates the ground state pairing gap as a function of the on-site interaction. When the evolution time following the quench is sufficiently long, the pairing gap typically approaches a finite value, suggesting that the system relaxes to a steady state. For large $U_f$, the pairing gap of the steady state is close to that in the ground state of the quenched Hamiltonian. For small $U_f$, the steady state pairing gap and that of the ground state at $U_f$ are noticeably different. In comparison, we demonstrate in Fig.~\ref{fig:evolutiondelta1} the quench dynamics where the system is initially in a TSF state. Different from the previous case, strong oscillations of $|\Delta(t)|$ are observed in the quenched state when $U_f$ is deep in the conventional superfluid regime. We observe no significant damping in the oscillations within
the longest evolution time
scale that we can numerically implement.

\begin{figure}
\includegraphics[width=1.0\linewidth]{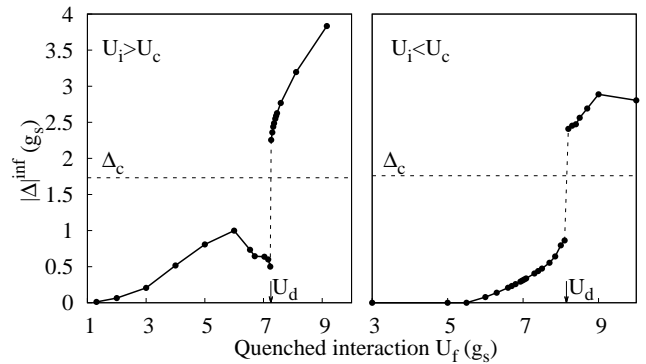}
\caption{Typical pairing gap in the long-time limit $|\Delta|^{inf}$ as a function of quenched interaction $U_f$ for a conventional initial state (left panel) and a topological initial state (right panel). The critical value $\Delta_c$ is plotted as the dashed lines in both panels.}\label{fig:decideud}
\end{figure}

Despite the difference in the detailed dynamics, for both of these two cases, we may numerically identify a critical final interaction $U_d$. As the quench parameter $U_f$ crosses $U_d$, the pairing gap $|\Delta|$ in the long-time limit features a jump. This is demonstrated in Fig.~\ref{fig:decideud}, where we plot the typical pairing gap $|\Delta|^{inf}$ in the long-time limit as a function of $U_f$. Here, $|\Delta|^{inf}$ is the asymptotic value of the pairing gap if the system approaches a steady state in the long-time limit; and is the time average of pairing gap in many periods in the case of strong oscillations. We associate this critical parameter $U_d$ with a dynamical phase transition. For a non-topological initial state, $U_d$ is numerically close but not equal to the first-order phase boundary in the ground state; while for a topological initial state, $U_d$ lies in the conventional superfluid regime, whose deviation from the topological phase boundary is much clearer. We have further checked
numerically that $U_d$ is not sensitive to the initial interaction $U_i$.

A particularly interesting observation here is that in either case, for $U_f<U_d$, $|\Delta|^{inf}$ is smaller than the critical value $\Delta_c$ that lies on the boundary of the topological phase transition of the initial state. In contrast, when $U_f>U_d$, $|\Delta|^{inf}$ is larger than $\Delta_c$. This strongly suggests that the dynamical phase transition is also a topological one.

\emph{Edge states in dynamical processes}.--
As the quenched state in the long-time limit is quite different from the ground state of the quenched Hamiltonian, the characterization of the topology of the quenched state is in general a highly non-trivial problem. Here, we propose to invoke the bulk-edge correspondence and examine the existence of edge states in the long-time limit~\cite{goldman13}. In heterostructure nanowires, a widely studied detection method for the Majorana edge state is by coupling the nanowire to a normal metal lead and measuring the tunneling conductance at zero bias. In the presence of a Majorana edge state, the resulting conductance is quantized to $2e^2/h$~\cite{law}. Here, we borrow this idea to theoretically study the existence of edges states in the quench dynamics. In particular, we examined the topology of the quenched state in the long-time limit. In practice, the transport measurement on cold atoms has been realized in recent experiments~\cite{stadler}, suggesting the possibility of carrying out similar measurements in
the future.

\begin{figure}
\includegraphics[width=1.0\linewidth]{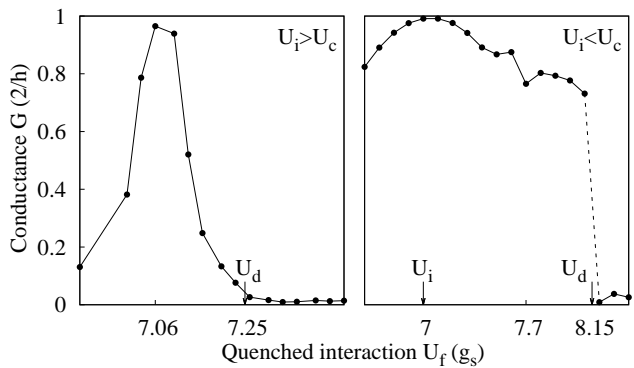}
\caption{The zero-bias differential conductance $G$ as a function of the quenched interaction $U_f$. In the calculation we set $g_V=g_s$ and $g_l=10g_s$. [Left panel] The initial state is SF at $U_i=7.85g_s$ and $n=1.19$, in correspondence to Fig.~\ref{fig:evolutiondelta3}. [Right panel] The initial state is TSF at $U_i=7g_s$ and $n=1.25$, in correspondence to Fig.~\ref{fig:evolutiondelta1}.}\label{fig:topoorder}
\end{figure}

We consider a spin-orbit coupled Fermi gas loaded into a lattice of $L$ sites, labeled as $0$ to $L-1$, and with open boundaries. A non-interacting Fermi gas in a half-infinite chain serves as a reservoir (lead), and is coupled to the edge site of the lattice gas labeled $0$. The Hamiltonian of the reservoir is then: $\hat H_{res} = -g_l \sum_{\sigma,j=-\infty}^{-2}(\hat c^\dag_{j\sigma} \hat c_{j+1,\sigma} +h.c.)$, with the coupling Hamiltonian:
$\hat H_{V} = g_V \sum_{\sigma} (\hat c^\dag_{-1,\sigma} \hat c_{0\sigma} +h.c.)$, where $\hat c_{0\sigma}$ and $c^\dag_{-1,\sigma}$ are field operators at the end of the interacting and noninteracting lattice respectively. The full Hamiltonian is then $\hat H_{tot}(t)=\hat H_{eff}(t) +\hat H_{res}+\hat H_{V}$.

Setting the chemical potential of the reservoir to $\mu_R$ while noticing that the gap of the superfluid is always symmetric with respect to the Fermi surface, we may define the zero-bias differential conductance at the junction as:
\begin{equation}\label{defdiffcond}
 G(t)= \left( -2 g_V\textbf{Im}\sum_\sigma  \frac{d}{d\mu_R} \langle \hat c^\dag_{-1,\sigma}(t) \hat c_{0,\sigma}(t) \rangle \right)_{\mu_R=0}.
\end{equation}
Here, the evolution of field operators $\hat c^\dag_{-1,\sigma}(t)$ and $\hat c_{0,\sigma}(t)$ is governed by $\hat H_{tot}(t)$, where the system and the reservoir are decoupled, and $\langle\cdots\rangle$ denotes the expectation value with respect to the initial state. We calculate $G(t)$ by the numerical operator method~\cite{supp,pei12,pei13a,pei13b}, and take the limit $L\to \infty$ before $t\to \infty$. Typically, we find that $G(t)$ relaxes to $G$ very fast.

Fig.~\ref{fig:topoorder} shows the zero bias conductance as a function of quenched interaction $U_f$, where the unit of conductance is $2/h$~\cite{stadler,unit}. When the initial state is a normal superfluid (the left panel), the conductance drops continuously to zero at the dynamical critical point $U_d$. When the initial state is a topological superfluid (the right panel), the conductance drops abruptly from a value close to $2/h$ to approximately zero at $U_d$. In either case, the behaviors of the conductance are qualitatively different when $U_f$ is on different sides of $U_d$. For $U_f<U_d$, the conductance is finite and has a peak which may touch $2/h$ at an appropriate $U_f$. While for $U_f>U_d$, the conductance is always close to zero as $U_f$ varies. These results strongly suggest the existence of edge states for quenches with $U_f<U_d$, regardless of the initial state. Thus, $U_d$ marks the onset of a dynamical topological phase transition. Typically, the dynamical critical point $U_d$ is close to
the equilibrium phase boundary $U_c$: in the left panel of Fig.~\ref{fig:topoorder}, $U_c\approx U_d= 7.25g_s$; while in the right panel they are $U_d= 8.15g_s$ and $U_c\approx 7.6g_s$.

Although the existence of edge states in the quench is not sensitive to the initial condition, the conductance in Fig.~\ref{fig:topoorder} behaves differently for $U_i<U_c$ and $U_i>U_c$. For $U_i<U_c$, $G$ reaches the quantized value at $U_f=U_i$ (no quench), and is close to $2/h$ in a wide range of $U_f$ near $U_i$, suggesting that the edge modes in quenched states are well protected before $U_f$ crosses the critical point. In contrast, for $U_i>U_c$, the emergence of edge states is accompanied by the crossing of the critical $U_c$ of $U_f$ in the quenched Hamiltonian. As the bulk gap in the ground state closes at $U_c$, the inevitable excitations in the quenched state make the signature of the edge states less obvious.

\emph{Conclusions}.--
In summary, we find a dynamical phase transition in cold fermions with synthetic spin-orbit coupling under an interaction quench. This transition is characterized by an abrupt change of asymptotic pairing gap, which also crosses the boundary distinguishing the topological and the trivial ground states. The edge state analysis strongly supports that this dynamical transition is a topological one, with the topological properties of the quenched state sensitively relying on the quench parameters. With the recent achievement of synthetic spin-orbit coupling and the high controllability of cold atomic gases, the dynamical topological phase transitions discussed here may be probed in future experiments.

\emph{Acknowledgement}.--
This work is supported by NFRP (2011CB921200, 2011CBA00200), NNSF (60921091), NSFC (11304280,11105134,11374283,11374266), the Fundamental Research Funds for the Central Universities (WK2470000006), and the Zhejiang Provincial Natural Science Foundation under Grant No. R6110175.

\clearpage

\appendix

\section{Supplementary material}
\subsection{Differential equations}
\label{app:diff}

In this section, we derive the equations of motion for the expectation values of the operators. As an example, we consider the expectation value of $\hat{\tau}_k$ in the Schr\"{o}dinger's picture:
\begin{equation}
 \tau_k(t)= \langle \psi(t)| \hat \tau_k |\psi(t)\rangle,
\end{equation}
where $|\psi(t)\rangle$ denotes the wave function at time $t$. The Schr\"odinger's equation gives:
\begin{equation}
i \frac{d |\psi(t)\rangle}{dt} = \hat H_{eff}(t) |\psi(t)\rangle.
\end{equation}
The equation of motion for $\tau_k(t)$ is then:
\begin{equation}
-i \frac{d \tau_k(t)}{dt}= \langle \psi(t)| \left[\hat H_{eff}(t), \hat \tau_k \right] |\psi(t)\rangle.
\end{equation}
The commutator on the right-hand side consists of six possible operators: $\{\hat \nu_{k\uparrow},\hat \nu_{k\downarrow},\hat n_{k\uparrow},\hat n_{k\downarrow},\hat \tau_k,\hat \sigma_k \}$. By calculating the derivative of their expectation values to $|\psi(t)\rangle$, we get a set of closed-equations for the expectation values of the operators:
\begin{equation}\label{diffequationgroup}
 \begin{split}
 -i \frac{d \nu_{k \uparrow} }{dt} = & 2\epsilon_{k\uparrow} \nu_{k\uparrow} + \Delta^* \sigma_{-k} - \Delta^* \sigma_k + \alpha_k \tau_{-k} + \alpha_k\tau_k \\
-i  \frac{d \nu_{k \downarrow} }{dt} = & 2\epsilon_{k\downarrow} \nu_{k\downarrow} + \Delta^* \sigma^*_{k} - \Delta^* \sigma^*_{-k} +  \alpha_k \tau_{-k} + \alpha_k\tau_k \\
-i \frac{d n_{k\uparrow}}{dt} = & - \Delta^* \tau_k^* + \Delta \tau_k -\alpha_k \sigma_k^* -\alpha_k \sigma_k \\
-i \frac{d n_{k\downarrow}}{dt} = & -\Delta^* \tau_{-k}^* + \Delta \tau_{-k} + \alpha_k \sigma^*_k + \alpha_k \sigma_k \\
-i \frac{d\tau_k}{dt} = & (\epsilon_{k\uparrow}+\epsilon_{k\downarrow}) \tau_k + \Delta^* n_{-k\downarrow} - \alpha_k \nu_{k\downarrow} + \Delta^* n_{k\uparrow} \\ & - \alpha_k \nu_{k\uparrow} - \Delta^* \\
-i \frac{d\sigma_k}{dt} = & (\epsilon_{k\uparrow}-\epsilon_{k\downarrow}) \sigma_k - \Delta^* \nu^*_{k\downarrow} - \Delta \nu_{k\uparrow} - \alpha_k n_{k\downarrow} \\ & + \alpha_k n_{k\uparrow}.
 \end{split}
\end{equation}
The pairing gap $\Delta(t)$ in Eq.~(\ref{diffequationgroup}) can be expressed as
\begin{equation}\label{defdeltaauxiliary}
\Delta(t) = U_f\sum_{k} \tau_k^{\ast}(t),
\end{equation}
which can be solved self-consistently from Eq.~(\ref{diffequationgroup}) and Eq.~(\ref{defdeltaauxiliary}).

\subsection{Calculation of tunneling conductances}
\label{app:method}

To calculate the tunneling conductance, we employ the numerical operator method first introduced in Ref.~\cite{pei12} for the real time Kondo problem, which was then extended to time-dependent systems~\cite{pei13a} and topological superconductors~\cite{pei13b}.

The full Hamiltonian describing the reservoir and the superfluid is expressed as
\begin{equation}
 \hat H_{tot}(t) = \hat H_{eff}(t) + \hat H_{res} + \hat H_V.
\end{equation}
We define the synonyms $\hat d_{j\sigma 1}=\hat c^\dag_{j\sigma}$ and $\hat d_{j\sigma 0} = \hat c_{j\sigma}$ and re-express the conductance as
\begin{equation}\label{defdiffcondapp}
 G(t)= \left( -2 g_V\textbf{Im}\sum_\sigma  \frac{d}{d\mu_R} \langle \hat d_{-1,\sigma, 1}(t) \hat d_{0,\sigma, 0}(t) \rangle \right)_{\mu_R=0}.
\end{equation}
We first calculate the operators $\hat d_{-1,\sigma, 1}(t)$ and $\hat d_{0,\sigma, 0}(t)$ in the Heisenberg picture defined as
\begin{equation}\label{defheisenbergoperator}
\begin{split}
 \hat d_{\alpha} (t)= & \lim_{\tau\to 0} e^{i\hat H_{tot}(t_0) \tau} \cdots e^{i\hat H_{tot} \left(t_{N-1} \right) \tau} \hat d_{\alpha} \\ & \times e^{-i\hat H_{tot} \left(t_{N-1} \right) \tau} \cdots e^{-i\hat H_{tot}(t_0) \tau},
\end{split}
 \end{equation}
where $\alpha$ is the abbreviation of $(j,\sigma, s)$ ($s=0,1$), $\hat H_{tot}$ is the time-dependent Hamiltonian, $N$ is the total number of steps, $t_0=0$ is the initial time, and the intermediate time $t_n=n\tau$, where the time step $\tau=t/N$.

The calculation of Eq.~(\ref{defheisenbergoperator}) is divided into $N$ steps with the time evolving from $t_{N-1}$ to $t_0$. In the {\it n}-th step, we undress the pair of operators $e^{i\hat H_{tot} \left(t_{N-n} \right) \tau} $ and $e^{-i\hat H_{tot} \left(t_{N-n} \right) \tau} $ from $\hat d_\alpha$. Since $\hat H_{tot}$ is quadratic, in principle, the intermediate results at each step can be written as
\begin{equation}\label{heisenbergsolution}
 \hat {\tilde d}_{\alpha} (t_n) = \sum_{\alpha'} W_{\alpha,\alpha'}(t_n) \hat d_{\alpha'},
\end{equation}
with $\hat {\tilde d}_{\alpha} (t_n) $ defined as
\begin{equation}\label{defintermediate}
\begin{split}
 \hat {\tilde d}_{\alpha} (t_n) = & e^{i\hat H_{tot}(t_{N-n}) \tau} \cdots e^{i\hat H_{tot} \left(t_{N-1} \right) \tau} \hat d_{\alpha} \\ & \times e^{-i\hat H_{tot} \left(t_{N-1} \right) \tau} \cdots e^{-i\hat H_{tot}(t_{N-n}) \tau},
\end{split}
 \end{equation}
 and $W_{\alpha,\alpha'}(t_n)$ denoting the propagator. At $n=N$, we get
\begin{equation}
\hat d_{\alpha} (t)=\sum_{\alpha'} W_{\alpha,\alpha'}(t_N) \hat d_{\alpha'}.
\end{equation}

To calculate the propagator $W_{\alpha,\alpha'}(t_N)$, we derive an iterative relation. According to Eq.~(\ref{defintermediate}), we have
\begin{equation}
\hat {\tilde d}_{\alpha} (t_n) =e^{i\hat H_{tot}(t_{N-n}) \tau} \hat {\tilde d}_{\alpha} (t_{n-1}) e^{-i\hat H_{tot}(t_{N-n}) \tau} .
\end{equation}
Substituting Eq.~(\ref{heisenbergsolution}) into the equation above, we find
\begin{equation}\label{iterativerelation}
\begin{split}
& \sum_{\alpha'} W_{\alpha,\alpha'}(t_n) \hat d_{\alpha'} \\ &
= \sum_{\alpha'} W_{\alpha,\alpha'}(t_{n-1})  \bigg( \hat d_{\alpha'}+ i\tau [\hat H_{tot}(t_{N-n}), \hat d_{\alpha'}] \\ & + \frac{(i\tau)^2}{2} [\hat H_{tot}(t_{N-n}),[\hat H_{tot}(t_{N-n}),\hat d_{\alpha'}]] + O(\tau^3) \bigg).
\end{split}
\end{equation}
The calculation of these commutators is straightforward with the results
\begin{equation}\label{commutatorexp}
[\hat H_{tot}(t_j), \hat d_{\alpha}]= \sum_{\beta} G_{\alpha,\beta}(t_j) \hat d_{\beta}.
\end{equation}
Substituting Eq.~(\ref{commutatorexp}) into Eq.~(\ref{iterativerelation}) and comparing the left and the right sides, we obtain the iterative relation
\begin{equation}\label{witerativerelation}
\begin{split}
 W_{\alpha,\alpha'}(t_n)= & W_{\alpha,\alpha'}(t_{n-1}) +i\tau \sum_{\beta} W_{\alpha,\beta}(t_{n-1}) G_{\beta,\alpha'}(t_{N-n}) \\ & -\frac{\tau^2}{2} \sum_{\beta,\beta'} W_{\alpha,\beta}(t_{n-1}) G_{\beta,\beta'}(t_{N-n}) G_{\beta',\alpha'}(t_{N-n})  .
\end{split}
\end{equation}
Here we drop terms of the order $O(\tau^3)$. In principle, the error caused by the discretization of time can be made arbitrarily small by letting $\tau\to 0$. In practice, we adaptively choose $\tau$ to make the discretization error negligible.

According to Eq.~(\ref{witerativerelation}), we start from $n=0$ when $W_{\alpha,\alpha'}(t_0)= \delta_{\alpha,\alpha'}$, and iteratively work out $W_{\alpha,\alpha'}(t_n)$ until $t_N=t$. This approach typically works well, due to the critical fact that the matrix $G_{\alpha,\beta}$ has only very few non-zero elements, such that the number of non-zero $W_{\alpha,\beta}$ which needs to be kept at each step increases slowly according to Eq.~(\ref{witerativerelation}). However, when $N$ is very large, which is inevitable since we must choose a small $\tau$ and at the same time a large $t$ to obtain the asymptotic limit of $G(t)$, there will be too many non-zero $W_{\alpha,\beta}$ which cannot be simultaneously kept. To work around this, we apply a truncation scheme and keep only a fixed number of non-zero propagators (the number is denoted by $M$) with the largest magnitudes at each step. This truncation scheme is necessary for obtaining the steady limit of $G(t)$. We decide the value of $M$ adaptively, i.e.,
set an original $M$ and
increase it until the desired precision is obtained. The logic in doing so is that the truncation error goes to zero in the limit $M\to \infty$.

Substituting the expression of $\hat d_{\alpha}(t)$ into Eq.~(\ref{defdiffcondapp}), we have
\begin{equation}
\begin{split}
 G(t)= & -2 g_V\textbf{Im}\sum_{\sigma\beta\beta'}  W_{-1\sigma 1,\beta}(t) W_{0\sigma 0,\beta'}(t) \\ & \times \left( \frac{d}{d\mu_R} \langle \hat d_{\beta} \hat d_{\beta'} \rangle \right)_{\mu_R=0}.
\end{split}
\end{equation}
Next we calculate the derivative of correlation functions. The correlation functions of the initial state are non-zero only if the two field operators are both in the superfluid or both in the reservoir, since the superfluid and the reservoir are decoupled at the initial time. Considering that the initial state of the superfluid located on sites from $0$ to $L-1$ is independent of $\mu_R$, we immediately get
\begin{equation}
 \frac{d}{d\mu_R} \langle \hat d_{j\sigma s} \hat d_{j'\sigma' s'} \rangle =  0,
\end{equation}
as $j,j' \geq 0$. As $j,j'< 0$, we notice that the reservoir is in the ground state of a free Fermi gas at the chemical potential $\mu_R$, and get
\begin{eqnarray}
\left\{ \begin{array}{c} \langle \hat d_{j\sigma 1} \hat d_{j'\sigma' 0} \rangle = \delta_{\sigma,\sigma'} \displaystyle \frac{\sin((j-j')\theta )}{\pi(j-j')} \\ \langle \hat d_{j\sigma 0} \hat d_{j'\sigma' 1} \rangle = \delta_{\sigma,\sigma'}\left( \delta_{j,j'}-\displaystyle \frac{\sin((j-j')\theta )}{\pi(j-j')} \right) \end{array} \right.
\end{eqnarray}
with $\theta=\arccos(-\mu_R/2g_l)$ and $g_l$ the hopping between two neighbor sites of the reservoir. Finally we find
\begin{eqnarray}
\begin{split}
\left(\displaystyle\frac{d}{d\mu_R} \langle \hat d_{j\sigma s} \hat d_{j'\sigma' s'} \rangle\right)_{\mu_R=0} =& \left[ (1-\delta_{s,s'})(2\delta_{s,1}-1) \right] \delta_{\sigma,\sigma'} \\ & \times \frac{\cos \left( (j-j')\displaystyle\frac{\pi}{2} \right)}{2g_l \pi}.
\end{split}
\end{eqnarray}
\end{document}